\documentclass[a4paper,11pt]{article}
\pdfoutput=1 

\usepackage{jinstpub} 

\title{\boldmath Latest R\&D news and beam test performance of the highly granular SiW-ECAL technological prototype for the ILC.}

\author[a]{A. Irles\note{Corresponding author.} on behalf of the CALICE collaboration}

\affiliation[a]{Laboratoire de l'Acc\'el\'erateur Lin\'eaire, Centre Scientifique d'Orsay, Universit\'e de Paris-Sud XI, CNRS/IN2P3, F-91898 Orsay Cedex, France}

\emailAdd{irles@lal.in2p3.fr}

\abstract{
  High precision physics at future colliders as the International Linear Collider (ILC) require unprecedented high precision in the determination
  of the energy of final state particles.
  The needed precision will be achieved thanks to the Particle Flow algorithms (PF) which require highly granular and hermetic calorimeters systems.
  The physical proof of concept of the PF was performed in the previous campaign of beam tests of physic prototypes within the CALICE collaboration.
  One of these prototypes was the physics prototype of the Silicon-Tungsten Electromagnetic Calorimeter (SiW-ECAL)
  for the ILC. 
  In this document we present the latest news on R\&D of the next generation prototype,
  the technological prototype with fully embedded very front-end (VFE) electronics,
  of the SiW-ECAL.
  Special emphasis is given to the presentation and discussion of the first results from the beam test done at DESY in June 2017.
  The physics program for such beam test consisted in the calibration and commissioning of the current set of available SiW ECAL modules;
  the test of performance of individual slabs under 1T magnetic fields; and the study of electromagnetic showers events.
}

\keywords{Calorimeter methods, calorimeters, Si and pad detectors}

\proceeding{Calorimetry for the High Energy Frontier 2017\\
  October 2-6, 2017\\
  Lyon, France}

\begin{document}
\maketitle
\flushbottom

\section{Introduction}

The International Linear Collider, ILC~\cite{Behnke:2013xla}, is an accelerator based particle physics project
which will provide collisions of polarized beams made up of $e^{+}e^{-}$ with center-of-mass energies ($c.m.e$) of 250 GeV - 1 TeV.
These collisions will be studied by two multipurpose detectors:
the International Large Detector (ILD) and the Silicon Detector (SiD)~\cite{Behnke:2013lya}.
The accomplishment of the ambitious physics program~\cite{Baer:2013cma,Fujii:2017vwa} of the ILC requires
unprecedented precision in the energy determination of final states.
To meet these required precision levels, the detectors will be based on the Particle Flow (PF) techniques~\cite{Brient:2002gh,Morgunov:2004ed}.
These techniques rely on single particle separation in the full detector volume to choose the best information available
to measure the energy of the final state objects (i.e. measuring the charged particles momentum at tracking devices better than in the calorimeters).
Therefore, PF algorithms require highly granular and compact calorimeter systems featuring minimum dead material (high hermeticity).
The R\&D of highly granular calorimeters for future linear colliders is conducted within the CALICE collaboration.
For further information about PF and CALICE R\&D we refer the reader to reference~\cite{Sefkow:2015hna} and references therein.

In this document we will focus in the description of the silicon-tungsten electromagnetic calorimeter, SiW-ECAL, technological prototype
and its performance in beam test. The design and R\&D of this prototype is conducted by CALICE and it is oriented at the baseline design of the ILD ECAL.
The ILD ECAL is a sampling calorimeter of 24 $X_{0}$ of thickness (in the barrel region) and it
uses silicon (Si) as active material and tungsten (W) as absorber material.
The combination of Si and W for the construction of the detector allows
the construction of a very compact calorimeter made up of compact active layers with small cell size (high granularity) in the transverse and longitudinal planes.
It will consist of an alveolar structure of carbon fiber into which the slabs made up of tungsten
plates and active sensors will be inserted. The very-front-end (VFE) electronics will be
embedded in the slabs.
The silicon sensors will be segmented
in squared cells of 5x5 mm, featuring a total of $\sim 100$ million channels for the ECAL of the ILD.
To reduce overall power consumption, the ILD ECAL will exploit the special bunch structure
foreseen for the ILC: the $e^{+}e^{-}$ bunchs trains will arrive within
acquisition windows of $\sim$ 1-2 ms width separated by $\sim$ 200 ms. During the idle time, $\sim99\%$ of the time, the bias currents of the electronics will be shut down.
This technique is usually denominated power pulsing. In addition, as the PF techniques demands minimum dead material in the detector,
the design of the ILD foresees the calorimeters (hadronic and electromagnetic) to be placed inside the magnetic coil that provides magnetic fields of 3.5 T.

\section{The SiW-ECAL engineering prototype.}

The first SiW-ECAL prototype was the so called SiW-ECAL physics prototype.
It was successfully tested at DESY, FNAL and CERN running together with another CALICE prototype,
the analogue hadronic calorimeter AHCAL, delivering the proof of concept of PF calorimetry.
For the physics prototype, the VFE was placed outside the active area with no particular constraints in power consumption.
Published results proving the good performance of the technology and the PF can be found in references~\cite{Adloff:2011ha,Anduze:2008hq,Adloff:2008aa,Adloff:2010xj,CALICE:2011aa,Bilki:2014uep}.

The new generation prototype is called the SiW-ECAL technological prototype  and it addresses the main technological challenges: compactness,
power consumption reduction through power pulsing and VFE inside the detector close to real ILD conditions.
It will also provide data allowing deep studies of the PF performance and input to tune the Monte Carlo programs.

The base unit of such technological prototype is the Active Sensor Units or ASU which is the entity of sensors, thin PCB (printed circuit boards) and ASICs (application-specific integrated circuits).
An individual ASU has a lateral dimension of 18x18 cm$^{2}$ and has glued onto it 4 silicon wafers (currently with a thickness of 320 $\mu$m).
The ASUs are equipped
further with 16 ASICs for the read out and features 1024 square pads, 64 per ASIC, of 5x5 mm$^{2}$ (the physics prototype featured squared pads of 10x10 mm$^{2}$).
The readout layers of the SiW-ECAL consist of a chain of ASUs and an interface card
to a data acquisition system (DAQ) at the beginning of the layer.
This interface card also carries services as power connectors,
test output pins, connectors for signal injection, etc. 
Currently, the technological prototype layers are built with a version of the PCB called FEV11 with
16 SKIROC~\cite{Callier:2011zz} (Silicon pin Kalorimeter Integrated ReadOut Chip) ASICs version 2
in BGA packages mounted on top of it. The SKIROC ASIC consists of 64 channels comprising each a low noise charge preamplifier of variable gain followed by two lines:
a fast line for the trigger decision and a slow line for dual gain charge measurement.
Finally, a Wilkinson type analogue to digital converter fabricates the digitized charge deposition that can be readout.
Once one channel is triggered, the ASIC reads out all 64 channels adding a bit of information to tag them as triggered or not triggered.
The information is stored in 15 cell deep physical switched capacitor array (SCA).
This autotrigger capability is mandatory for the ILC case since the accelerator will not provide a central global trigger.
A key feature of the SKIROC ASICs is that they can be power pulsed to meet ILC power consumption requirements.
A new version, 2a, has been produced and will be used to equip new layers currently in production.
The design of the subsequent chain of the data acquisition (DAQ) is described in~\cite{Gastaldi:2014vaa}. 
The whole system is controlled by the Calicoes and the Pyrame DAQ software version 3~~\cite{pyrame1,pyrame2}.

\begin{figure}[!t]
  \centering
  \includegraphics[width=5in]{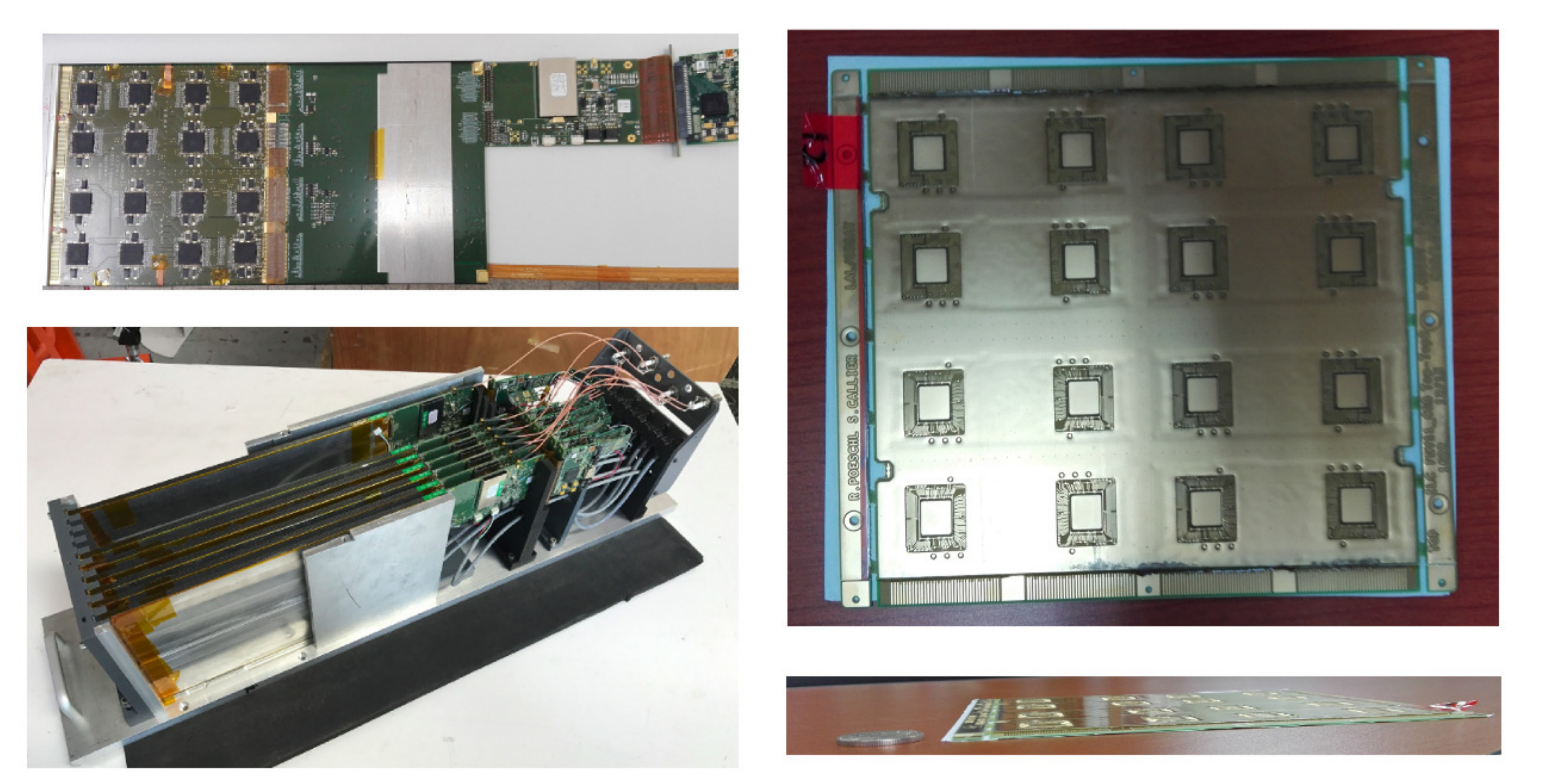}
  \caption{Leftmost upper figure: open single SLAB with FEV11 ASU, 16 SKIROC, interface card and DIF visible; the silicon sensors are glued to the PCB in the other side. Leftmost lower figure: picture of the technological prototype with 7 single short layers inside a mechanical aluminum structure designed for beam tests. Rightmost photographs show the FEV11\_COB: the upper figure correspond to a picture taken from the top; the lower one corresponds to a lateral picture.}
  \label{proto}
\end{figure}

Figure \ref{proto}, leftmost upper plot, shows a picture of a full equipped short slab with FEV11 ASU. 
These PCBs still don't meet current requirements
for the ILD in terms of thickness (1.2 mm). The FEV11 thickness is 1.6 alone and 2.7 mm including the ASICs.
There are ongoing R\&D activities in an alternative PCB design in which the ASICs
will be directly placed on board of the PCB in dedicated cavities. The ASICS will be in semiconductor packaging and wire bonded to the PCB.
These PCBs are denominated COB for "chip on board".
A small sample of FEV11\_COB boards with thickness of 1.2 mm have been produced (see figure \ref{proto} rightmost photographs)
and is planned to be added to the prototype and tested in beam tests conditions.

\section{Performance on positron beam test.}

The prototype tested in beam in June 2017 consisted of 7 layers: see figure \ref{proto} leftmost lower photograph.
A dedicated and comprehensive commissioning process was followed before going to the beam test.
The commissioning included the definition of trigger threshold values or the list noisy channels to be masked.
In the first layer, $\sim 40\%$ of channels were masked due to a damaged Si wafer. In all the other, only the 6-7\% of channels were masked except in the last one were this number grew up to the 16\% due to a faulty ASIC. The number of masked channels may be drastically reduced by setting individual threshold settings instead of global trigger threshold values for each channel on an ASIC. This possibility will become available with the next version of SKIROC ASIC.

The detector was exposed to a positron beam in the DESY test beam area (line 24).
The beam test line at DESY provides continuous positron beams in the energy range of 1 to 6 GeV with
rates of the order of the KHz (with maximum of $\sim 3$ KHz). In addition, DESY gives acces to a bore 1 T solenoid in the beam area. 
The detector was running in power pulsing mode without any extra active cooling system.
By means of an external pulse generator we defined the length of the acquisition window to be
3.7 ms at a frequency of 5 Hz. 

The physics program of the beam test can be summarized in the following points:

\begin{itemize}
\item Commissioning and calibration without absorber using 3 GeV positrons acting as minimum ionizing particle (MIPs);
\item magnetic field tests up to 1 T;
\item response to electrons with fully equipped detector, i.e. sensitive parts {\it and} W absorber.
\end{itemize}

\subsection{Calibration runs.}

\begin{figure}[!t]
  \centering
  \begin{tabular}{ll}
    \includegraphics[width=2.6in]{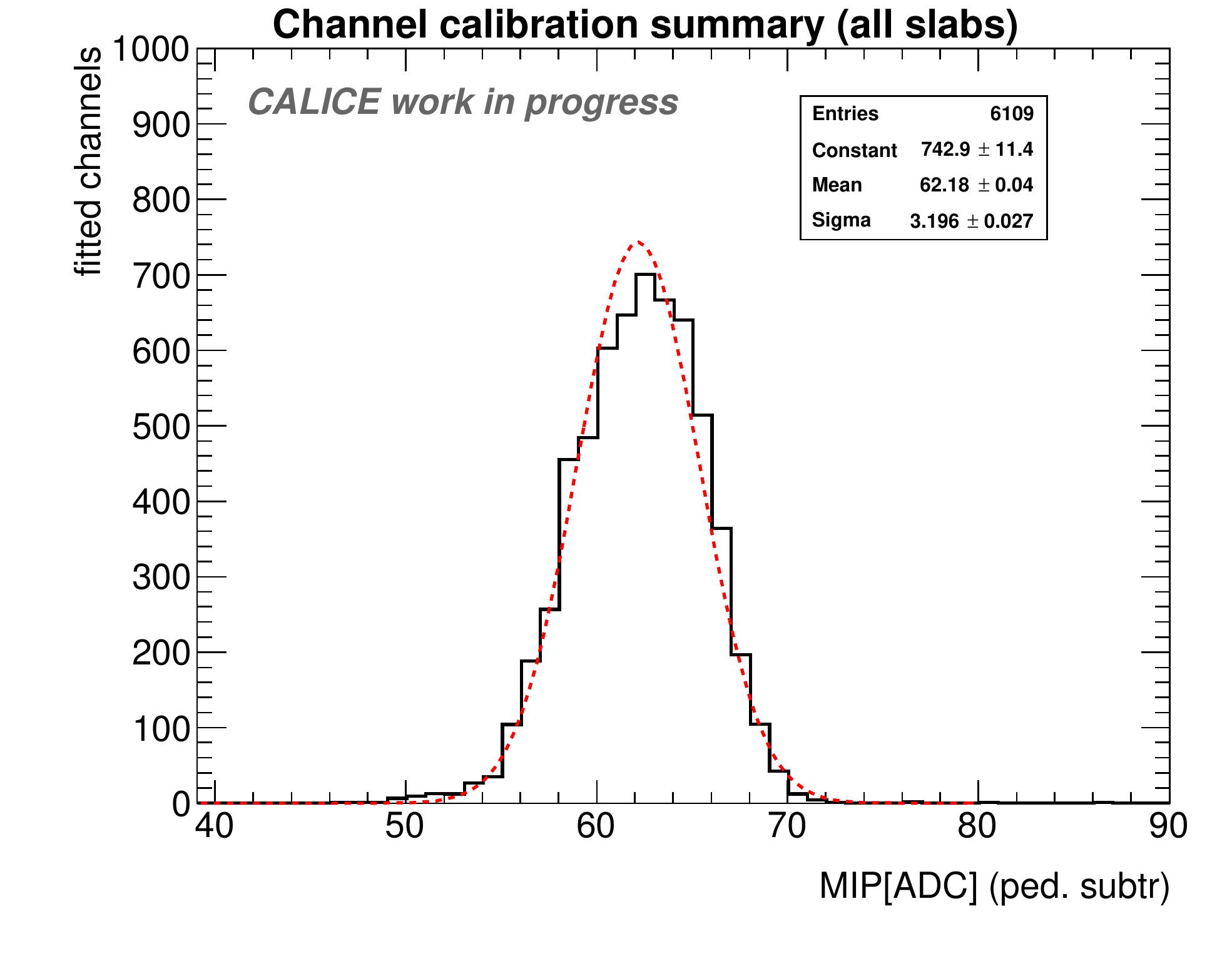} & \includegraphics[width=2.6in]{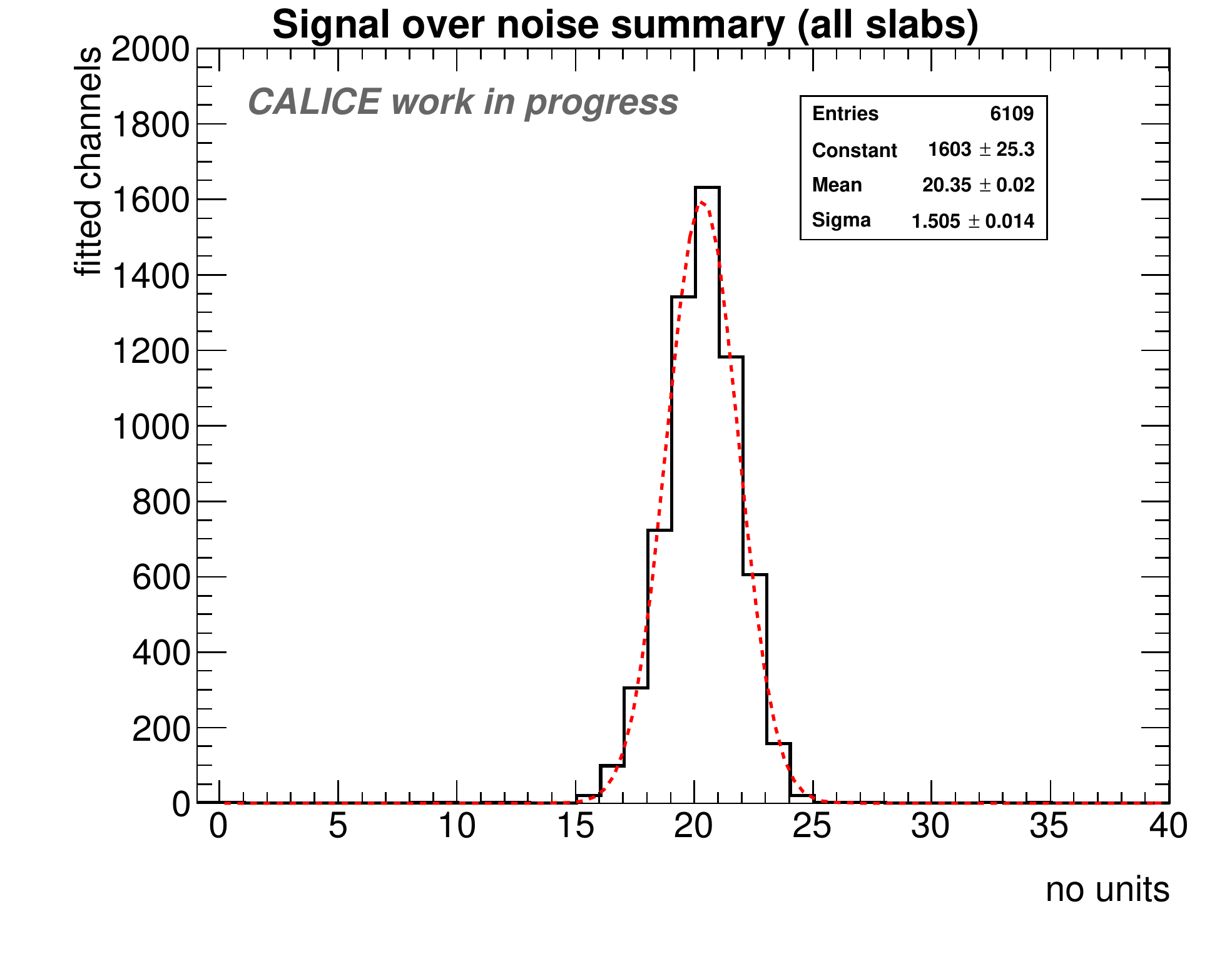}
  \end{tabular}
\caption{Result of the MIP position calculation and signal over noise calculation for all calibrated cells.}
\label{mipandSN}
\end{figure}
  
The main calibration was realized 
by directing the 3 GeV positron beam on 81 positions equally distributed over the surface of the detector.
These data were used for pedestal estimation and energy calibration.
Calibration and pedestal analysis was done for all single layers not requiring track reconstruction.
We calculated the pedestal position for every channel and SCA by fitting the distribution of non triggered hits with a Gaussian function.
Afterwards, we subtracted these values to the distribution of triggered hits and fit the resulting distributions to a Landau function convoluted by a Gaussian.
The most-probable-value of the convoluted function is taken as the MIP value.
We have obtained a raw energy calibration spread of the 5\% among all cells with the 98\% of all available cells being fitted. Results are summarized in figure \ref{mipandSN}, leftmost plot.

The signal-over-noise ratio, defined as the ratio between the most-probable-value of
the Landau-gauss function fit to the data (pedestal subtracted) and the pedestal width
(calculated as the standard deviation of the Gaussian distribution fitted to the data), was estimated.
The average value for all channels and slabs is 20.4.
Results are summarized in figure \ref{mipandSN}, rightmost plot.

\begin{figure}[!t]
  \centering
    \begin{tabular}{ll}
      \includegraphics[width=2.6in]{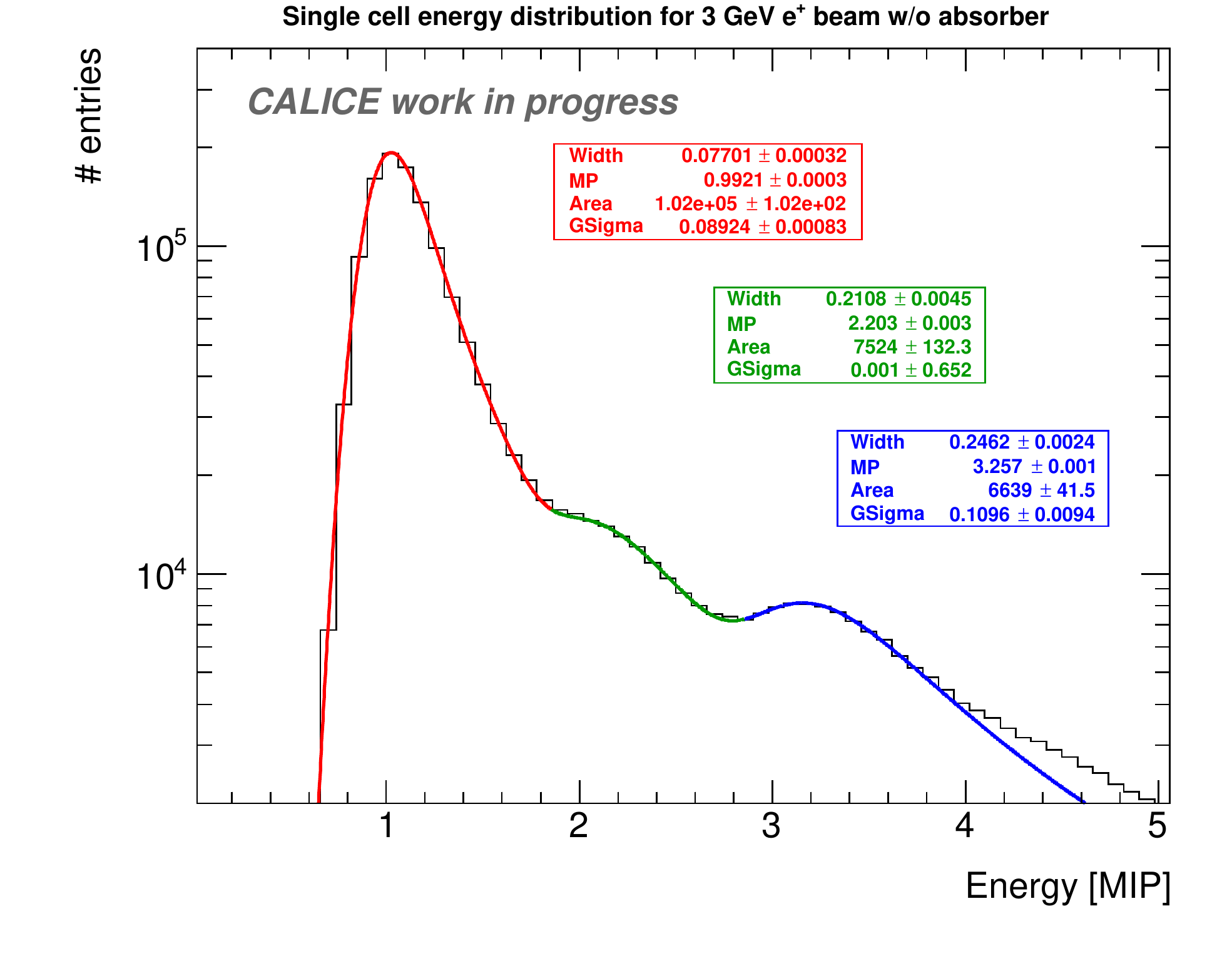} & \includegraphics[width=2.6in]{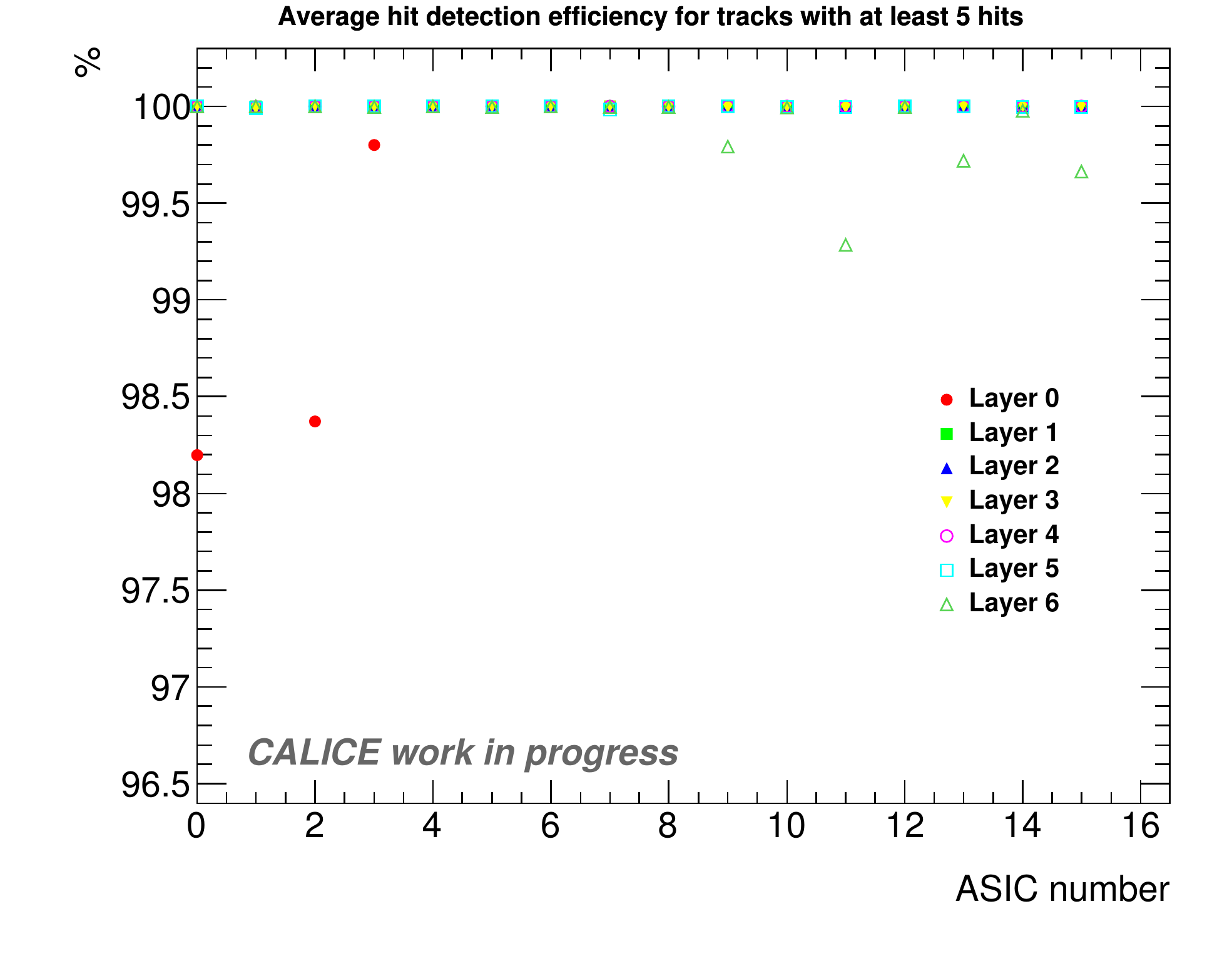} \\
    \end{tabular}
    \caption{Left: the single cell energy distribution (for all calibrated cells) for 3 GeV positron tracks acting as MIPs. Right: hit efficiency for all layers and ASICs in high purity samples of tracks of MIP-like acting particles.}
\label{miplog}
\end{figure}

After pedestal subtraction, calibration and track reconstruction we could finalize the MIP calibration by selecting tracks that cross the detector parallel to its normal.
The results are shown in figure \ref{miplog} where, in the leftmost plot, single cell the energy distribution for MIPs is shown for all calibrated cells. 
The distribution reveals the presence of a second and third peak due to events involving multiple particles crossing the detector.
In the rightmost plot, we
summarize the results of hit detection efficiency in tracks made of MIP acting particles. To evaluate the efficiency we define a high purity sample of
events of positron traversing the detector perpendicularly by selecting
tracks with at least 5 layers (of 7 possible) with a hit in exactly the same cell. Afterwards we check if the other layers have or not a hit in the same cell. Finally, we repeat this for all layers and cells and show the result for all ASICs and layers. Except few exceptions, the efficiency is compatible with $100\%$.
Lower efficiencies may be related to some channels having effective high thresholds for the triggers.
This can be improved with the next generation of SKIROC (the 2a) which allow for threshold optimization of single channels.
To avoid the confusion of the concept of inefficiency and blindness of the detector due to saturation of the DAQ
(for example if a noisy channels fills up the memory before the physical signal) we constraint the analysis to events that are stored before the last but one cell in the SCA.

Finally, a calibration run with the beam hitting the slabs under an angle of $\sim 45^{0}$ was done.
The purpose of this run was to prove that the MIP position scales with a factor $\sqrt{2}$
due to the larger path to be crossed by the positron in the Si wafer.
Preliminary results show a perfect agreement with the expected results.

\subsection{Magnetic field tests.}

For this test, a special PVC structure was
designed and produced to support the slab.

The purpose of the test was twofold: first to prove that the DAQ, all electronic devices and the mechanical slab itself were able
to handle strong magnetic fields; second to prove the stability of performance during these tests.
We took several runs, with 0, 0.5 and 1 T magnetic fields with and without 3 GeV positron beam.
We observed that the pedestals position is independent of the magnetic field (within the 1 per mille). The MIP position was increased, in average,
by 3\% for 1 T and 1.5\% for 0.5 T with respect with the 0 T case.
This level of increase is expected since the positron traversing the magnetic field hit the slab with a deflecting angle,
increasing then the path of the particle through the wafer. More detailed studies and simulation comparisons are to come.

\subsection{Response to electromagnetic showers.}

\begin{figure}[!t]
  \centering
  \begin{tabular}{ll}
    \includegraphics[width=2.6in]{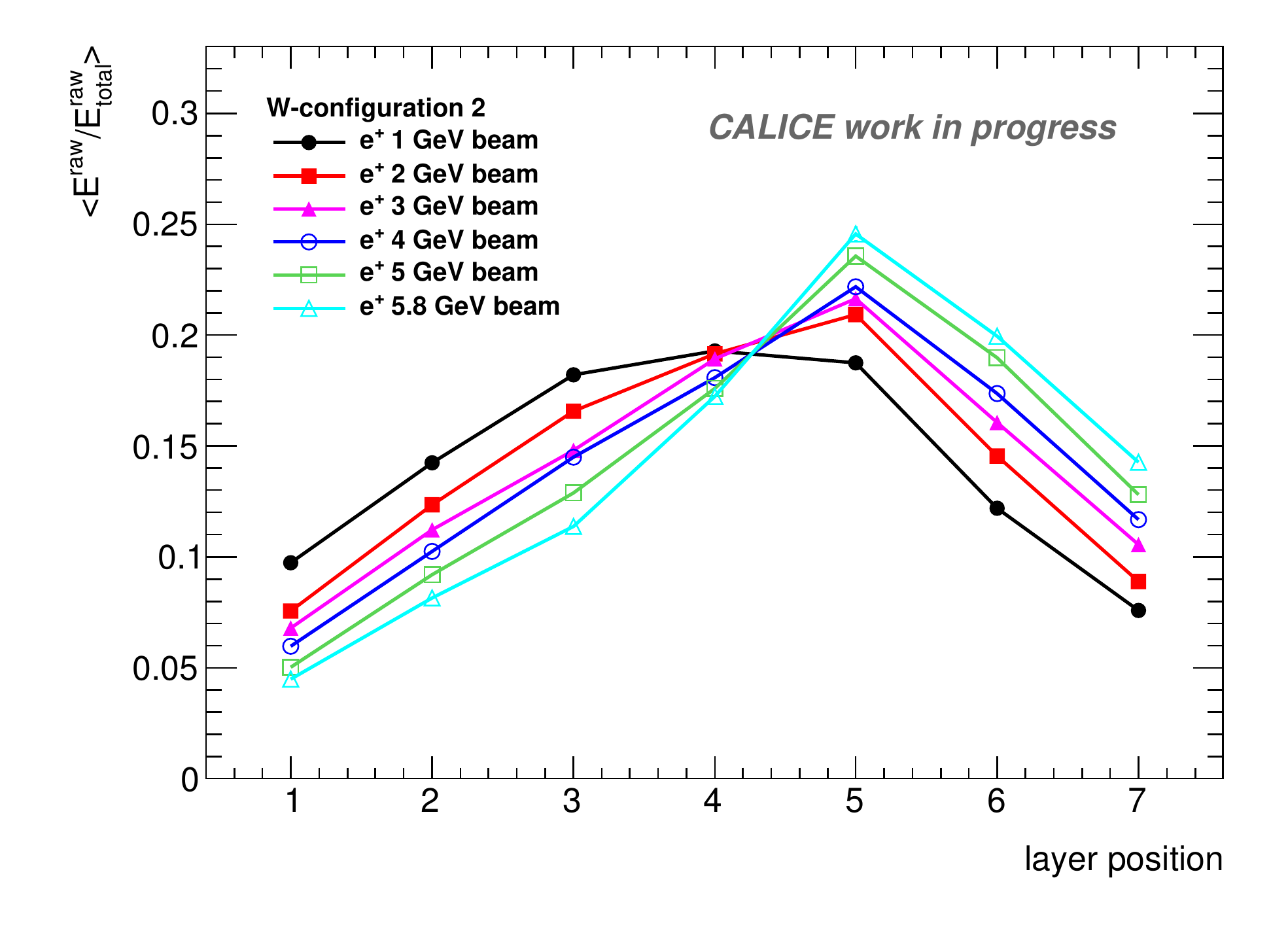} & \includegraphics[width=2.6in]{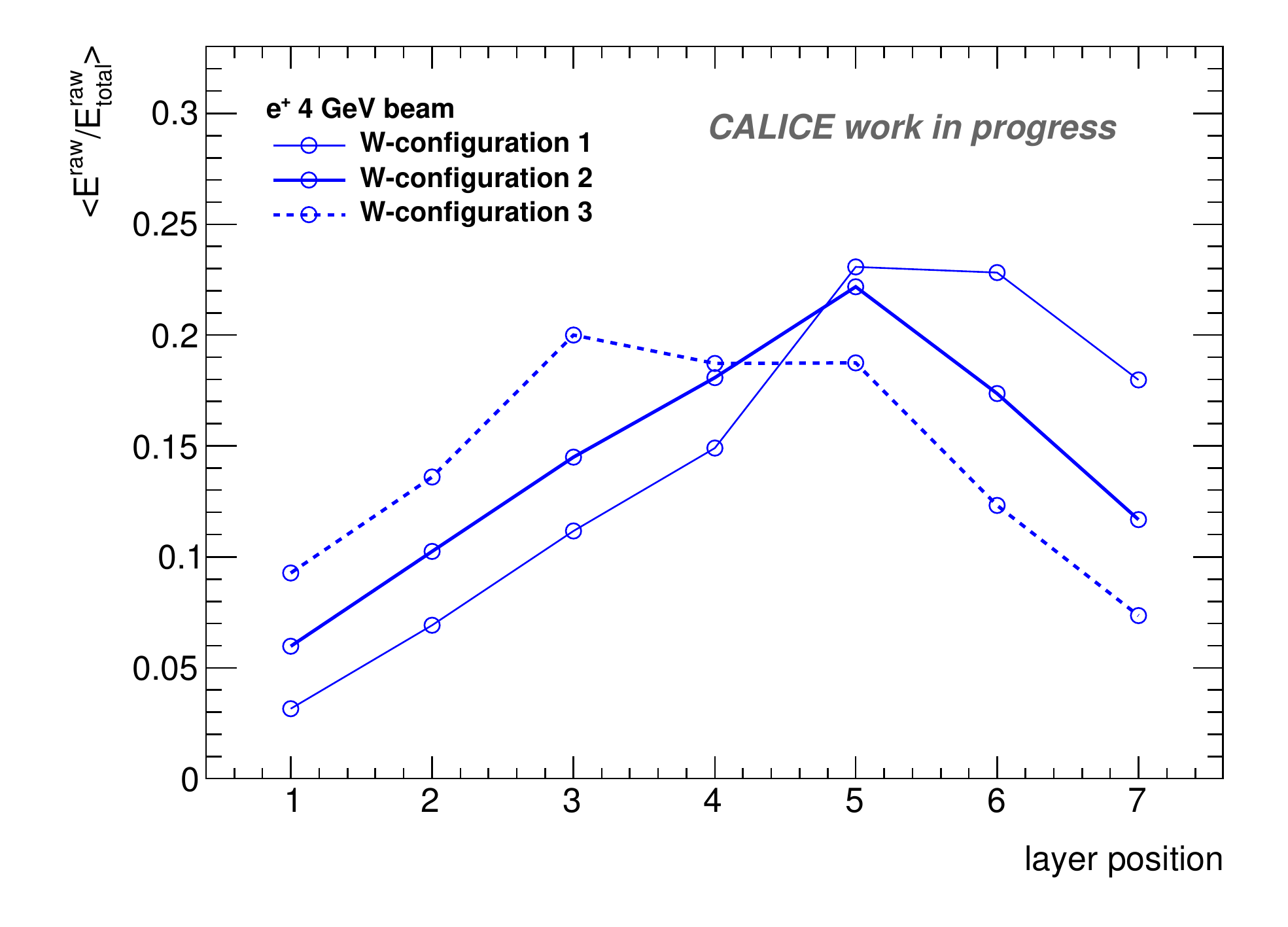} 
  \end{tabular}
  \caption{Raw electromagnetic shower profiles for different tungsten configurations and several energies of the beam. In the x-axis, we show the layer number. In the y-axis, the averaged fraction of {\it energy} (sum of ADC in all triggered cells in a event, considering only events where all layers had at least a hit) measured in every layer. }
\label{showers}
\end{figure}

The purpose of the test was to study the interaction of positrons with the absorber material resulting in
electromagnetic showers.
We inserted W plates of different thicknesses between the sensitive layers and we performed
a scan of energies of the positron beam: 1-5.8 GeV.
We tested the response of the detector with three different configurations of the W repartition.
The accumulated amount of tungsten, in radiation length units, $X_{0}$, in front of each of the modules is:
\begin{itemize}
\item W-configuration 1: $0.6,1.2,1.8,2.4,3.6,4.8$ and $6.6~X_{0}$
\item W-configuration 2: $1.2,1.8,2.4,3.6,4.8,6.6$ and $8.4~X_{0}$
\item W-configuration 3: $1.8,2.4,3.6,4.8,6.6,8.4$ and $10.2~X_{0}$
\end{itemize}
Preliminary results of the raw electromagnetic shower profiles are shown in figure \ref{showers}
for several beam energies in the W-configuration 2 (left) and for 4 GeV beam energy and the three different W-configurations (right). This first approach to the data looks promising but further studies and comparisons with simulations are needed.

\acknowledgments

This project has received funding from the European Union{\textquotesingle}s Horizon 2020 Research and Innovation program under Grant Agreement no. 654168.
This work was supported by the P2IO LabEx (ANR-10-LABX-0038), excellence project HIGHTEC,
in the framework {\textquotesingle}Investissements d{\textquotesingle}Avenir{\textquotesingle}
(ANR-11-IDEX-0003-01) managed by the French National Research Agency (ANR).
The research leading to these results has received funding from the People Programme (Marie
Curie Actions) of the European Union{\textquotesingle}s Seventh Framework Programme (FP7/2007-2013)
under REA grant agreement, PCOFUND-GA-2013-609102, through the PRESTIGE
programme coordinated by Campus France.
The measurements leading to these results have been performed at the Test Beam Facility at DESY Hamburg (Germany), a member of the Helmholtz Association (HGF).


\end{document}